\input harvmac
\noblackbox
\newif\ifdraft

\catcode`\@=11
\newif\iffrontpage
\newif\ifxxx
\xxxtrue

\newif\ifad
\adtrue
\adfalse

\parindent0pt

\def\{{\lbrace}
\def\}{\rbrace}

\def\sss{\scriptscriptstyle}


\def\Box#1{\mathop{\mkern0.5\thinmuskip
           \vbox{\hrule\hbox{\vrule\hskip#1\vrule height#1 width 0pt\vrule}
           \hrule}\mkern0.5\thinmuskip}}
\def\box{\displaystyle{\Box{7pt}}}   
%
%


%
\def\abstract#1{
\vskip.5in\vfil\centerline
{\bf Abstract}\penalty1000
{{\smallskip\ifx\answ\bigans\leftskip 2pc \rightskip 2pc
\else\leftskip 5pc \rightskip 5pc\fi
\noindent\abstractfont \baselineskip=12pt
{#1} \smallskip}}
\penalty-1000}

\def\sss{\scriptscriptstyle}

\vbadness=10000

\parindent10pt

%
\lref\AGMOO{O.~Aharony, S.S.~Gubser, J.~Maldacena, H.~Ooguri and Y.~Oz,
``Large N field theories, string theory and gravity,''
Phys.\ Rep.\ 323 (2000) 183 [hep-th/9905111].}
\lref\BS{
S.~Deser,
``Closed form effective conformal anomaly actions in $d\geq 4$,''
Phys.\ Lett.\  {B479}, 315 (2000)
[hep-th/9911129].}
\lref\BONORA{L.~Bonora, P.~Pasti and M.~Bregola, 
``Weyl cocylcles,'' Class.\ Quant.\ Grav.\ 3, 635 (1986).}
\lref\VANDRIEL{O.~Coussaert, M.~Henneaux and P.~van Driel, ``The asymptotic
dynamics of three-dimensional Einstein gravity with a negative
cosmological constant,'' Class. Quant. Gravity (1996) 296 [gr-qc/9506019]; 
M.~Banados, K.~Bautier, O.~Coussaert, M.~Henneaux and M.~Ortiz,
Phys.\ Rev.\  {D58}, 085020 (1998)
[hep-th/9805165];
M.~Henneaux, L.~Maoz and A.~Schwimmer,  ``Asymptotic Dynamics and Asymptotic 
Symmetries of Three-Dimensional Extended AdS Supergravity,''
Annals of Physics 282 (2000) 31 [hep-th/9910013].}
\lref\DS{S.~Deser and A.~Schwimmer,
``Geometric classification of conformal anomalies in arbitrary dimensions,''
Phys.\ Lett.\ {B309} (1993) 279
[hep-th/9302047].}
\lref\FG{C. Fefferman and R. Graham, ``Conformal Invariants,'' 
Ast\`erisque, hors s\'erie, 1995, p.95.}
\lref\wittenone{E.~Witten,
``Anti-de Sitter space and holography,''
Adv.\ Theor.\ Math.\ Phys.\ {2} (1998) 253
[hep-th/9802150].}
\lref\tensor{S.de Haro, E.Bautier}
\lref\ISTY{
C.~Imbimbo, A.~Schwimmer, S.~Theisen and S.~Yankielowicz,
``Diffeomorphisms and holographic anomalies,''
Class.\ Quant.\ Grav.\  {17}, 1129 (2000)
[hep-th/9910267].}
\lref\Pen{R.~Penrose and W.~Rindler, ``Spinors and Spacetime'', 
vol 2, CUP 1986, chapter 9.}
\lref\APTY{O.~Aharony, J.~Pawelczyk, S.~Theisen and S.~Yankielowicz,
``A note on anomalies in the AdS/CFT correspondence,''
Phys.\ Rev.\ {D60} (1999) 066001
[hep-th/9901134].}
\lref\HS{M.~Henningson and K.~Skenderis,
``The holographic Weyl anomaly,''
JHEP {07} (1998) 023
[hep-th/9806087].}
\lref\BPB{L.~Bonora, P.~Pasti and M.~Bregola, ``Weyl cocycles'', 
Class. Quantum. Grav. 3 (1986) 635.}
\lref\CFL{A.~Cappelli, D.~Friedan and J.I.~Latorre,
``C theorem and spectral representation,''
Nucl.\ Phys.\ {B352} (1991) 616.}
\lref\BH{J.D.~Brown and M.~Henneaux,
``Central Charges In The Canonical Realization Of Asymptotic Symmetries: 
An Example From Three-Dimensional Gravity,''
Commun.\ Math.\ Phys.\ {104} (1986) 207.}
\lref\Strings{
http://strings99.aei-potsdam.mpg.de/cgi-bin/viewit.cgi?speaker=Theisen}
\lref\Cardy{J.L.~Cardy,
``Is There A C Theorem In Four-Dimensions?,''
Phys.\ Lett.\ {B215} (1988) 749.}
\lref\BGNNO{S.~Nojiri and S.D.~Odintsov,
``On the conformal anomaly from higher derivative gravity in 
AdS/CFT  correspondence,'' hep-th/9903033;
M.~Blau, K.S.~Narain and E.~Gava,
``On subleading contributions to the AdS/CFT trace anomaly,''
JHEP {09} (1999) 018 [hep-th/9904179].}
\lref\Solo{S.~de Haro, S.~N.~Solodukhin and K.~Skenderis,
``Holographic reconstruction of spacetime and renormalization 
in the  $AdS/CFT$ correspondence,'' hep-th/0002230.}
\lref\Eng{K.~Bautier, F.~Englert, M.~Rooman and P.~Spindel,
Phys.\ Lett.\  {B479}, 291 (2000) [hep-th/0002156].}

\Title{\vbox{
\rightline{\vbox{\baselineskip12pt
\hbox{AEI-2000-045}
\hbox{hep-th/0008082}}}}}
{Diffeomorphisms, 
Anomalies and the}
\vskip-1cm
{\titlefont\centerline{Fefferman-Graham Ambiguity
\footnote{$^{\scriptscriptstyle*}$}{\sevenrm 
Partially supported by GIF, 
the German-Israeli Foundation for Scientific Research,
by the European Commission TMR programmes ERBFMRX-CT96-0045 and 
FMRX-CT96-0012, 
Minerva Foundation,
the Center for Basic Interactions of the Israeli Academy of Sciences.
}}}
\vskip 0.3cm
\centerline{ A.~Schwimmer$^a$ and 
S.~Theisen$^b$ }
\vskip 0.6cm
\centerline{$^a$ \it Department of Physics of the Complex Systems, 
Weizmann Institute, Rehovot 76100, Israel} 
\vskip.2cm
\centerline{$^b$ \it Max-Planck-Institut f\"ur Gravitationsphysik, 
Albert-Einstein-Institut, 14476 Golm, Germany}
\vskip 0.0cm

\abstract{Using the Weyl transfomations induced by diffeomorphisms
we set up a cohomological problem for the Fefferman-Graham coefficients.
The cohomologically nontrivial solutions remove the ambiguity and give
the nonlocal terms in the effective action 
responsible for the trace anomalies.}
\Date{\vbox{\hbox{\sl {August 2000}}
}}
\goodbreak

\parskip=4pt plus 15pt minus 1pt
\baselineskip=15pt plus 2pt minus 1pt

\newsec{Introduction}

The AdS/CFT correspondence offers remarkable insight into
nonperturbative phenomena in gauge theories \AGMOO.
Many of the proposed tests of the correspondence rely on the symmetry
algebras being isomorphic.

Among the tests going beyond the mapping of 
the algebraic structure the correct mapping of the trace anomalies is one
of the most impressive \wittenone,\HS,\APTY. 

On the CFT side, when the theory is put in a general gravitational
background, the effective action contains specific nonlocal terms with local
Weyl variations. The trace anomalies are produced by these terms.

On the supergravity side the correspondence involves a classical calculation:
one solves the equation of motion using the metric at the boundary as initial
condition. The action evaluated for this classical solution gives the
effective  action in terms of the boundary metric.  An anomaly appears in a 
classical calculation due to the apparently 
infrared logarithmically divergent terms obtained when the action is 
evaluated  with the classical solution.  In this treatment, though 
the anomaly is correctly reproduced, the origin of the nonlocal
terms in the action  responsible for it  is not clear.

In this note we will try to answer this question.
In \ISTY\ the relation between the Weyl invariance in even 
dimension and
a certain subgroup of diffeomorphisms in odd
dimension (called the ``PBH
transformation'') was studied. It was shown that  these transformations 
determine universally the structure
of the Euler (type A) trace anomaly.  
In the present note we study further the structure of the PBH transformations,
in particular their implications 
for the structure of the Fefferman-Graham (FG in the following)
expansion \FG\ and its
physical interpretation where  we will find the origin of the nonlocal terms.

The FG expansion deals with the solution of the equations
of motions for an  AdS action. The metric is expanded
in powers of the radial variable and the coefficients of
the expansion determined iteratively as local expressions
in terms of the boundary metric. As shown in \FG\ in odd dimension $d+1$ ,
where  $d=2n$ the expansion has the following feature:
 
(i) the first $n-1$ coefficients are determined uniquely as local
$2n$-dimensional diffeo-covariant expressions in terms of the boundary
metric;
  
(ii) the expansion breaks down at order $n$; a logarithmic dependence
on the radial variable should be introduced beginning at this order;

(iii) the $n$-th order coefficient is not determined completely
by the equations of motion: only its trace and covariant divergence
are determined iteratively as local and diffeo-invariant expressions
(the ``FG ambiguity'').

Combining the PBH transformations and (iii) we set up a cohomological
problem involving the expansion coefficients.
We show that exactly the $n$-th order coefficient has a nontrivial
cohomology and we classify it.
Then  features (i) and (ii) listed above are shown to be consequences of
the cohomological structure uncovered.
Using the solution of the cohomology problem the complete
expression for the $n$-th order coefficient
is obtained which is nonlocal.

The gravitational action needs boundary terms. The boundary action was 
evaluated in \Solo,\Eng\ in terms of the FG coefficients.
When the nonlocal terms are inserted into the boundary action
they provide the nonlocal terms in the effective action which
generate the anomalies.
Since the results are
algebraic, being  based on the behaviour of the coefficients
under the PBH transformations, we expect an analogous
procedure to hold also when stringy
corrections are included in the gravitational lagrangian.

Due to the nonchiral nature of the Weyl transformations 
it is extremely convenient to use dimensional regularization.
When $d$ is non-integer, the FG expansion gives  well-defined
local expressions for all the coefficients. The special features
happening at order $n$ mentioned above are signaled by the appearance
of poles in $ d-2n $. The way in which poles in dimensional
regularization are translated into cohomological features in even integer
dimension was studied in \DS. We summarize this procedure, tuned to the
present problem, in Section 2.

In Section 3 we set up the cohomological problem and we discuss
the consistency conditions involved.

In Section 4 we  classify the solutions of the cohomological problem
and their relation to the $n$-th order coefficient in the FG expansion.

The general conclusions we have reached
and possible implications 
and open questions are discussed in the last Section.  
The  discussion of the special features of the cohomology related to the Euler
anomaly is relegated to an Appendix. 

\newsec{The relation between Weyl cohomologies and dimensional regularization}

We start with the well understood case of the Weyl anomalies \DS.

Consider a conformal theory coupled to an external, c-number 
gravitational field $g_{ij}(x)$. The effective action, after integrating
out the fields of the conformal theory, is $W(g)$.

In an even dimension the anomaly 
can be formulated as a cohomological problem \BONORA, i.e.
searching for local Weyl variations which
cannot be obtained as the variation of a local action.
 
If we use dimensional
regularization, {\it i.e.} we consider the theory in dimension $d$, there
are two interrelated features which make the treatment  of anomalies
particularly simple:

1) In $d$ dimensions there are no nontrivial cohomologies:
there is always a local expression whose variation 
gives the anomaly in $2n$ dimensions. These local
expressions have, however,
a pole in $d-2n$. We distinguish between two situations:

i) The residue of the pole  vanishes exactly in
$d=2n$ due to a ``topological'' identity. This is the origin
of the so called ``type A anomaly''; we will continue to call 
more generally ``type A'' expressions in dimensional regularization those 
which have a pole and whose residue vanishes for $d=2n$ due to some special
identity. 

ii) The residue of the pole does not vanish  for $d=2n$. We will call this
situation ``Type B''. The effective action related to this case in $d=2n$
will have a scale.

2) In $d$ dimensions there are no anomalies, {\it i.e.} 
the effective action is exactly Weyl invariant.
It follows that the aforementioned local pole terms
will be accompanied by finite, nonlocal terms such that the sum is invariant.
As a consequence the variation of the pole term is equal (but opposite in sign)
to the variation of the nonlocal term (the anomaly).

We illustrate the above general discussion with two typical examples.

We start with ``type A''. In $d$ dimensions we expand the invariant
effective action
around $d=2n$. The leading term is local while the order $d-2n$ term
is nonlocal. The effective action has therefore a pole term with a
local residue which vanishes in $d=2n$ and a nonlocal finite term.
 The simplest example of this kind is in $d=2$  :
in $d=2$ the trace anomaly appears in the correlator of two energy-momentum 
tensors of the conformal field theory. Using Lorentz invariance, conservation
and tracelessness of the energy-momentum tensor, 
the structure of the correlator is fixed uniquely, up to an overall 
constant $a$ ($d=2-\epsilon$):

\eqn\polyy{\eqalign{
\langle T_{ij}(q)T_{kl}(-q)\rangle&=a\left\lbrace
-{(q_i q_j-\delta_{ij}q^2)(q_k q_l-\delta_{kl}q^2)\over q^{2+\epsilon}}
\right.\cr
&\qquad\left.+{\delta_{ij}\delta_{kl}q^2-\delta_{ik}\delta_{jl}q^2
+\delta_{ik}q_j q_l+\delta_{jl}q_i q_k-\delta_{ij}q_k q_l
-\delta_{kl}q_i q_j\over\epsilon}\right\rbrace\,.}}
This structure in flat space leads to an analogous one when the CFT is put
in a general background $g_{ij}$. The generating functional $W(g)$, after 
the CFT is integrated out, has the expansion:
\eqn\polyyy{ 
W(g)=-{1\over4}\int d^dx\sqrt{g}R{1\over\box^{1+{\epsilon\over2}}}R
+{1\over d-2}\int d^d x\sqrt{g}R\,,}
where $E_2=R $ is the two-dimensional Euler density. As usual
in dimensional regularization, the tensors are defined in $d$-dimensions.
$W(g)$ is Weyl invariant in $d$ dimensions to order $\epsilon$ as is easy
to verify. This can be derived directly by expanding a certain Weyl invariant
expression in $d$ dimensions, around $d=2$. The second term, with
the $0/0$ structure, signals the presence in $d=2$ of a nontrivial 
cohomology in the Weyl transformations.

Once we have the expansion
\polyyy\ we can discuss how to take the limit $d\rightarrow 2$:
we simply drop the second, local term which has the $0/0$ structure, and
therefore the limit is the first, finite, nonlocal term.

On the other hand, we can still use the local term if we 
are interested in the Weyl variation (anomaly) of the limit: 
since the total Weyl 
variation of \polyyy\ is $0$, the variation of the local term is equal and 
opposite in sign to the variation of the limit, both variations being finite.
Indeed the Weyl variation of the pole term gives $\sqrt g R\sigma$.

Summarizing the lessons learned from this simple example:

a) The  pole term in $d$ dimensions 
with a vanishing residue signals the presence of a nontrivial
cohomology in integer (even) dimensions.

b) If a certain variation of the pole term has a well-defined limit
then  the pole term indicates that in integer dimension we
should take a nonlocal expression which has the same variation.
The functional dependence of the nonlocal term can be inferred by 
``completing'' the pole term to a Weyl invariant expression in $d$ dimensions.

Now we discuss the ``type B'' case, again in the simplest
situation where it occurs, {\it i.e.} the second Weyl anomaly in $d=4$:

Considering a CFT  around $d=4$ in a general background, the correlator 
of two energy-momentum tensors gives rise to a term in the effective
action of the form:
\eqn\weyl{W_d(g)={1\over d-4}\int d^d x\sqrt{g}C^{(d)}_{ijkl}
\box^{{d-4\over2}}C^{(d)ijkl}\,,}
where $C^{(d)}$ is the Weyl tensor in $d$ dimensions ({\it c.f.} 
Appendix A). 
The expression
\weyl\ is Weyl invariant to order $d-4$.\foot{For recent proposals to
extend this expression to an exact one see \BS.}

The expression \weyl\ does not have a well-defined limit for
$d\rightarrow 4$. One should modify it by a ``subtraction'' :
\eqn\weylsub{W^{\rm (sub)}_d(g)=W_d(g)
-{\mu^{d-4}\over d-4}\int d^d x\sqrt{g}C^{(d)}_{ijkl}C^{(d)ijkl}\,.}
It is important to remark that the ``subtraction'' introduces a scale
$\mu$.
Now the limit can be taken, giving in $d=4$:
\eqn\weylfour{W_4(g)={1\over2}\int d^4 x\sqrt{g}C_{ijkl}
\log\left({\box\over\mu^2}\right)C^{ijkl}\,.}
We can rewrite $W_d$ as :
\eqn\split{\eqalign{
W_d(g) & =
{1\over2}\int d^d x\sqrt{g}C^{(d)}_{ijkl}\log\left({\box\over\mu^2}\right)
C^{(d)ijkl}+{\mu^{d-4}\over d-4}\int d^d x\sqrt{g}C^{(d)}_{ijkl}C^{(d)ijkl}\,.
}}
The expression \split\ is analogous to  \polyyy, {\it i.e.} we have a local
pole term and a nonlocal finite term, the finite term being the limit in the
integer dimension. Similarly to the ``type A'' situation the 
finite, local Weyl variation
of the nonlocal term can be calculated by using the fact that
$W_d$ is invariant and therefore the variation of $W_4$ is equal
and opposite in sign to the variation of the pole term.
In this particular case this gives $\sqrt g C_{ijkl} C^{ijkl}\sigma$.
 
The general conclusions we want to draw from this example for the general
``type B'' situation are as follows:

a) The presence of a pole term with a nonvanishing residue in
dimension $d$ indicates that the theory in integer dimension 
has to be modified by defining it after a ``subtraction''.

b) After the subtraction the relevant expression can be rewritten as
a sum of two terms, one finite and nonlocal and the other a local pole term
with nonvanishing residue. The limit is the nonlocal term.
 
c) The local variation can be calculated directly from the pole term.

We are now ready to use the above lessons for the study of the structure 
of the FG expansion.

\newsec {Consistency conditions of the FG coefficients}
Consider a manifold in $d+1$ dimensions with a boundary
which is topologically $S_d$.
Following Fefferman and Graham \FG, one can choose a set of coordinates
in which the $d+1$ dimensional metric has the form:
\eqn\ansatz{
ds^2=G_{\mu\nu} dX^\mu dX^\nu={l^2\over4}\left({d\rho\over\rho}\right)^2
+{1\over\rho}g_{ij}(x,\rho)dx^i dx^j\,.
}
Here $\mu,\nu=1,\dots,d+1$ and $i,j=1,\dots,d$. The coordinates are 
chosen such that $\rho=0$ corresponds to the boundary. We will assume
that $g_{ij}$ is regular at $\rho=0$,
$g_{ij}(x,\rho=0)$ being the boundary metric. 

We further assume that  $g_{ij}(x,\rho)$ 
has a power series expansion in the vicinity of $\rho=0$, {\it i.e.} we write 
\eqn\expansiong{
g_{ij}(x,\rho)=\sum_{n=0}^\infty g_{{\sss (n)}ij}(x)\rho^n\,.}
As shown by FG \FG, when the expansion \expansiong\ is inserted 
into the equations of motion following from the Einstein action with negative
cosmological constant in $d+1$ dimensions, one obtains a 
set of recursive equations for the $g_{{\sss (n)}ij}(x)$. We summarize
the features of the solutions as studied by FG:

1) In non-integer $d$ dimensions all the $g_{{\sss (n)}ij}$'s are
local, diffeo-covariant expressions which are uniquely 
determined in terms of the boundary metric.

2) The functions $g_{{\sss (n)}ij}$ have poles at $d=2n$ but their
covariant divergences $\nabla ^i g_{{\sss (n)}ij}$ and traces
$g_{\sss (0)}^{ij}g_{{\sss (n)}ij}$ are finite.

3) When $d$ is taken to $2n,\, n>1$, the expansion breaks down since
the equations become inconsistent. Starting
with the $\rho^n$ term, a $\log(\rho)$ dependence should be added but
even after adding these terms the equations of motion determine
only the covariant divergence and trace of $g_{{\sss(n)}ij}$.

We will study the structure of the $g_{{\sss (n)}ij}$, first directly
in $d=2n$. As first conditions abstracted from the study of FG
we will require that $g_{{\sss (n)}ij}$ has a local covariant divergence
and trace  which we will call $ V_{{\sss (n)}i}$ and $ S_{\sss (n)}$, 
respectively.

Next, we will use the transformation properties of $g_{{\sss (n)}ij}$
under Weyl transformations. In \ISTY\ it was shown that a certain subgroup
of the diffeomorphisms in $d+1$ dimensions
(called ``PBH transformations'') induces transformations
of $g_{{\sss (n)}ij}$ when the boundary metric undergoes a Weyl
transformation with parameter $\sigma(x)$. The transformation
has the general form:
\eqn\var{\delta g_{{\sss (n)}ij}= 2 \sigma (1-n) g_{{\sss (n)}ij}
+ A_{{\sss (n)}ij}(\sigma)\,,}
where $A_{{\sss (n)}ij}(\sigma)$ is a local, finite, covariant functional
of $g_{ij}$ \foot{When there is no danger of confusion, we will simply
denote by $g_{ij}$ the boundary metric $g_{{\sss (0)}ij}$.}
and $\sigma$, linear in the infinitesimal $\sigma $.
The PBH transformations determine, to a large extent,
$A_{{\sss (n)}ij}(\sigma)$ recursively but in this section we will not
use this information.

The structure we want to study is contained in the three local functionals 
$V_{{\sss (n)}i},\, S_{\sss (n)}, \,A_{{\sss (n)}ij}$.
Obviously, these functionals should fulfil some consistency conditions.  

We start with $ A_{{\sss (n) }ij}$: if we perform a second Weyl
variation  on \var\ the result should not
depend on the order since the Weyl group is abelian.
This gives a Wess-Zumino type condition for 
$A_{{\sss (n)}ij}$: 
\eqn\wza {2(1-n)\sigma_1 A_{{\sss(n)}ij}(\sigma_2)
+\delta_{\sigma_2}A_{{\sss(n)}ij}(\sigma_1)=
2(1-n)\sigma_2 A_{{\sss(n)}ij}(\sigma_1)
+\delta_{\sigma_1}A_{{\sss(n)}ij}(\sigma_2)\,.}
Following from the definitions of $V_{{\sss(n)}i}$ and $S_{\sss (n)}$
as covariant divergence and trace, respectively, of $g_{{\sss (n)}ij}$,
their Weyl variation is given in terms of $A_{{\sss(n)}ij}(\sigma)$.
Using \var\ and calculating explicitly the variations in $d=2n$ we
obtain:
\eqn\wzb{\delta S_{\sss(n)}=-2n\sigma S_{\sss(n)}
+g^{ij}A_{{\sss(n)}ij(\sigma)}\,,}
\eqn\wzc{\delta V_{{\sss(n)}i}=-2n\sigma V_{{\sss(n)}i}
+\nabla^j A_{{\sss(n)}ij}(\sigma)-S_{\sss(n)}\nabla_i\sigma\,.}
Now we can formulate our ``cohomological'' problem:
We search for triples of {\it local} 
functionals $ V_{{\sss (n)}i},\, S_{\sss (n)},\,
A_{{\sss (n)}ij}(\sigma)$ which satisfy the consistency conditions
\wza, \wzb, \wzc; clearly, any local functional $g_{{\sss (n)}ij}$,
by taking its covariant divergence, trace and Weyl variation
\var, generates such a triple. When a consistent triple does not have
a local generator, the  nonlocal $g_{{\sss(n)}ij}$ producing
the triple constitutes a nontrivial  solution. Of course, like
in all problems of this type, the nontrivial solutions are really equivalence
classes under the addition of local generators. In the above treatment we 
will consider all the quantities to be covariant under $d$-dimensional
diffeomorphisms.

We illustrate the above structure for the highly degenerate $n=1,\, d=2$ case.
The only possible $V_i$ and $S$ are $\partial_i R$ and $R$, respectively.

For $A_{ij}$ we have as possible candidates $g_{ij}R\sigma$,
$\nabla_i\nabla_j\sigma $ and $g_{ij}\box\sigma$.
The first expression does not fulfil \wza. The other two, 
after using \wzb\ and \wzc, give the following two consistent triples:
\eqn\triplea{\lbrace\partial_i R,2 R, 2 g_{ij}\box\sigma\rbrace\,,}
\eqn\tripleb{\lbrace 0,{1\over2}R,
g_{ij}\box\sigma-\nabla_i\nabla_j\sigma\rbrace\,.}
The first triple is trivial, being generated by $g_{ij}R$; the second
is a nontrivial solution corresponding to:
\eqn\po{\bar g_{{\sss (1)}ij} ={1\over{\sqrt g}}{\delta\over
{\delta g^{ij}}}W^P(g)\,,}
where $ W^P(g)$ is the  two-dimensional, nonlocal Polyakov action:
\eqn\podef{W^P(g)=-{1\over4}\int d^2 x\sqrt{g}R{1\over\box}R\,.}

Going back to our physical problem, the variation $A_{{\sss (1)}ij}$
is fixed uniquely by the  recurrence relations following from the PBH
transformations \ISTY\ to a linear combination of the ones appearing
in the above two solutions. Therefore, reintroducing the normalization
in terms of the AdS radius $l$ in $d=2$ we have the expression:
\eqn\sol{g_{{\sss (1)}ij}={l^2\over 2}[ g_{ij}R-2\bar g_{{\sss (1)}ij}]}
which satisfies $\delta_\sigma g_{{\sss(1)}ij}=l^2\nabla_i\nabla_j\sigma$, 
as it should \ISTY.
After this simple example we will study the general problem in the 
next section. 

\newsec{The general analysis of the cohomology problem}

As a first step we study how $A_{{\sss (n)}ij}$, which are solutions
of \wza, can be constructed. There are two basic mechanisms:

i) If a diffeo-invariant, dimensionless functional of $ g_{ij}$, 
$W(g)$, has a  Weyl variation for an infinitesimal Weyl parameter
$\sigma$, which can be expressed as an integral over 
a {\it local} density $D(g)$
\eqn\act{\delta_{\sigma} W(g)=\int d^{2n}x\sqrt g D(g)\sigma}
then $A_{{\sss (n)}ij}$, defined as 
\eqn\adef{A_{{\sss (n)}ij}={1 \over \sqrt g} {\delta \over {\delta g^{ij}}}
\delta_{\sigma} W(g)\,,}
fulfils \wza.
The proof is simple. Consider the commutator between the operators:
\eqn\com{\Bigl[{1\over\sqrt g}{\delta\over{\delta g^{ij}(x)}},
\delta_{\sigma}\Bigr]
=2(n-1)\sigma(x){1\over\sqrt{g}}{\delta\over{\delta g^{ij}(x)}}\,.}
Applying \com\ to $W(g)$ we obtain \var\ for a $ g_{{\sss (n)}ij}$
defined as:
\eqn\ggdef{g_{{\sss (n)}ij}= {1\over\sqrt g}{\delta\over{\delta g^{ij}}}W(g)\,.}
Therefore, $A_{\sss(n)}$ will fulfil \wza\ automatically and it is local
by assumption. Since, obviously, a local $W(g)$ will give a local
$g_{{\sss (n)}ij}$ \ggdef, the interesting solutions will reflect
nonlocal $W(g)$'s, {\it i.e.} the actions which generate the Weyl anomalies.
Then $A_{{\sss (n)}ij}$ is expressed in \adef\ directly in terms of the
anomaly $\sqrt g D(g)\sigma$.

Now $D(g)$ can be either $E_{2n}$, the Euler characteristic (``type A''
anomaly), or an expression which together with $\sqrt g$ is
locally Weyl invariant  (``type B'', whose number
increases with the dimension). We will distinguish the corresponding
variations by indexing them with $E$ and $B$, respectively.

We postpone the discussion of a subtlety related to the ``type A '' until
we introduce the other conditions and we proceed with the second mechanism:

ii) If a covariant local functional of $g_{ij}$, $F_{{\sss(n)}ij}$,
transforms homogeneously under a Weyl transformation:
\eqn\homo{\delta _{\sigma} F_{{\sss (n)}ij}= 2(1-n)\sigma F_{{\sss (n)}ij}}
then
\eqn\ssol{ A_{{\sss (n)}ij}\equiv F_{{\sss (n)}ij} \sigma}
satisfies \wza\ as can be easily verified.

Therefore the problem is reduced to the construction of
functionals which transform homogeneously. Such functionals 
are directly related to the $D(g)$ which give ``type B'' anomalies,  discussed
above. Each such functional $D^B (g)$ gives two homogeneously
transforming functionals:\hfill\break 
a) simply define:
\eqn\inv{ F_{{\sss (n)}ij}= g_{ij} D^B\,;}
b) for a given $D^B$ define:
\eqn\bdef{ B_{{\sss (n)}ij} \equiv {1\over {\sqrt g}} 
{\delta \over {\delta g^{ij}}} W^D}
with $W^D$:
\eqn\ddef{ W^D(g) \equiv \int d^{2n}x \sqrt g D^B(g)\,.}
In order to show that indeed $B_{{\sss(n)}ij}$ transforms homogeneously,
one simply applies the commutator \com\ to $W^D$, taking into account
that $W^D$ is Weyl invariant. Following from its definition,  
$B_{{\sss (n)}ij}$ has vanishing trace and covariant divergence since $W^D$
is invariant under Weyl transformations and under diffeomorphisms.

Now, after having a list of solutions of \wza, we have to study
which of the solutions give rise to nontrivial cohomologies.

For the solutions mentioned under i) we discuss now a special
feature related to the ``Type A'' - Euler one: the variation
$A_{{\sss (n)}ij}^E $ related to it by \adef\ can be obtained also
as the inhomogeneous term
in the variation of a $local$ $ \bar g_{ij}^E$. The proof based on an 
explicit construction of  $ \bar g_{ij}^E$
is presented in the Appendix. We conclude therefore that from the point
of view of \wza\ alone, $A_{{\sss (n)}ij}^E $ would be trivial.

However,
calculating the covariant divergence and trace corresponding to i)
we obtain the triples
\eqn\typea{\lbrace 0,-{1\over 2}E_{2n},A_{{\sss (n)}ij }^E\rbrace\,,}
\eqn\typeb{\lbrace 0,-{1 \over 2} D^B ,A_{{\sss (n)}ij }^B\rbrace\,.}
As triples, these are cohomologically nontrivial. The special feature
discussed in the Appendix allows us  to choose for the class \typea\ a
representative 
\eqn\homo {(\nabla^j\bar g_{(n)ij}^E, 
g^{ij}\bar g_{(n)ij}^E+{1\over 2}E_{2n},0)\,,}
i.e. transforming homogeneously under Weyl transformations. Since the trace
part of the triplet is an invariant made of Weyl tensors,
we can always choose a representative such that also the trace part
is zero; therefore all the nontrivial cohomological information resides
in the first term.
 
The nontrivial solutions constructed under ii)
are eliminated once we include
\wzb, \wzc. 
Indeed, inhomogeneous pieces in the Weyl transformation proportional to $D^B$
themselves are eliminated since the variation of the trace 
should obey:
\eqn\tracev{\delta_{\sigma} S=-2 n \sigma S +\sigma D^B\,.}
Taking $\sigma$ to be $x$ independent \tracev\ really measures the ordinary
dimension of $S$ and if $S$ has a well-defined dimension, \tracev\ cannot
hold.

The $ B_{{\sss (n)}ij}$ constructed above have zero trace and zero 
covariant divergence
as we remarked and
therefore, by \wzb, $S$ transforms homogeneously.
Eq.~\wzc\ becomes:
\eqn\covd{\delta_{\sigma}V_{{\sss(n)}i}
=-2n \sigma V_{{\sss(n)}i}+B_{{\sss(n)}ij}\nabla^j\sigma
-S_{(n)}\nabla_i\sigma\,.}
However, there are no local $V_{{\sss(n)}i}$ 
with the transformation properties given by \covd.
Therefore we are left with only the nontrivial triples constructed 
under i) which are in one-to-one correspondence with the Weyl anomalies.
Now we are ready to discuss the role of the nontrivial triples in the
FG expansion and their physical interpretation.

As a general method we will use the FG solutions in dimensional regularization
supplemented by the analysis of their variations in integer dimensions.
 
We start with the two-dimensional case already analysed directly at the end 
of Section 3. From the PBH transformation one obtains \ISTY:
\eqn\firstges{\eqalign{
g_{{\sss(1)}ij}&=
{l^2\over d-2}\Bigl[R_{ij}-{1\over 2(d-1)}R g_{{\sss(0)}ij}\Bigr]=
{l^2\over d-2}\Bigl[R_{ij}-{1\over 2}R g_{{\sss(0)}ij}\Bigr]+
{l^2\over 2(d-1)}R g_{{\sss(0)}ij}\,.}}
In the first term in \firstges\ we recognize ${X_{{\sss (1)}ij}\over d-2}$
defined in  (A.3),
signaling the presence of the nontrivial solution \typea. Therefore we can
directly write down the solution \sol\ for $g_{{\sss(1)}ij}$.

We proceed now to discuss the four-dimensional case where 
the special features related to the ``type B" solution first appear. 
We use again the solution of the PBH equations
calculated in  \ISTY:
\eqn\secondges{\eqalign{g_{{\sss(2)}ij}&
=c_1 l^4 C^{\sss (d)}_{ijkl}C^{{\sss (d)}ijkl}g_{{\sss(0)}ij}
+c_2 l^4 C^{\sss (d)}{iklm}C_{j}{}^{{\sss (d)}klm}\cr
&+{l^4\over d-4}\Bigl\lbrace -{1\over 8(d-1)}\nabla_i\nabla_j R
+{1\over 4(d-2)}\box R_{ij}
-{1\over 8(d-1)(d-2)}\box  R g_{{\sss(0)}ij}\cr
&-{1\over 2(d-2)}R^{kl}R_{ikjl}
+{d-4\over 2(d-2)^2}R_i{}^k R_{jk}
+{1\over (d-1)(d-2)^2}R R_{ij}\cr
&+{1\over 4(d-2)^2}R^{kl}R_{kl}g_{{\sss(0)}ij}
-{3d\over 16(d-1)^2 (d-2)^2}R^2 g_{{\sss(0)}ij}\Bigr\rbrace\,.}}
Here $C^{(d)}_{ijkl}$ is the Weyl-tensor in $d$ dimensions
which transforms homogeneously under Weyl transformations
({\it c.f.} Appendix  for the definition of $C^{(d)}$ and other
notation used below).
We can rewrite \secondges\ as follows:
\eqn\secsec{\eqalign{
&g_{{\sss(2)}ij}=\bar c_1 l^4 C^{(d)}_{ijkl}C^{(d)ijkl}g_{{\sss(0)}ij}
+\bar c_2 l^4 C^{(d)}_{iklm}C^{(d)}_{j}{}^{klm}\cr
&~-{l^4\over{16(d-3)}}{B_{{\sss(2)}ij}^{(d)}\over d-4} 
+a {l^4\over{16(d-3)}}\Bigl({X_{{\sss(2)}ij}\over d-4} 
-\bar g_{{\sss (2)}ij}^E\Bigr) +{l^4\over16(d-3)}
\Bigl(4(d-3)\bar R_{ik}\bar R_j^k-4 C_{ikjl}^{(d)}\bar R^{kl}\Bigr)}}
where $B_{{\sss (2)}ij}^{(d)}$ is given by \bdef,
$D^B$ being in this case the only ``type B" anomaly in $d=4$, {\it i.e.}:
\eqn\bano{ D^B_4 \equiv C_{ijkl}^{(d)}C^{{(d)}ijkl}}
and
\eqn\rdef{\bar R_{ij}\equiv{1\over d-2}\Bigl(R_{ij}
-{1\over{2(d-1)}}g_{ij}R\Bigr)\,.}
In \secsec\ the solution of the PBH recursion relation allows a general 
coefficient $a$; the solution of the equations of motion 
for the AdS action $S=\int d^{d+1}x\sqrt{G}(R(G)-2\Lambda)$ 
requires $a=1$ and $\bar c_1=\bar c_2=0$ \HS.
The nontrivial, nonlocal solutions in $d=4$ are signaled 
by the pole terms in \secsec.
The ${X_{{\sss (2)}ij}\over d-4}$
term indicates, in complete analogy to the $d=2$
case discussed before, the nonlocal type A term corresponding to
\typea\ with $n=2$.

The new feature is the appearance of the 
${B_{{\sss (2)}ij}^{(d)}\over d-4}$
pole term. Its residue doesn't vanish. Therefore, as we discussed in Section 2,
the limit $d\rightarrow 4$ cannot be taken: 
one needs a subtraction. The physical
motivation for the subtraction can be understood 
if we consider the Weyl variation
of the pole term. We obtain:
\eqn\bbvar{\delta_\sigma{B_{{\sss (2)}ij}^{(d)}\over d-4}=-2\sigma
{B_{{\sss (2)}ij}^{(d)}\over d-4}+A_{{\sss (2)}ij}^B
-\sigma B_{{\sss (2)}ij}\,,}
where $A_{{\sss (2)}ij}^B$ is given by \adef\ with the choice  $D^B_4$  for the
$D$ in \act. As we discussed above under i), 
$A_{{\sss (2)}ij}^B$ corresponds to
\typeb\ and has a well-defined nonlocal solution in $d=4$. On the other hand,
the second term $\sigma B_{{\sss (2)}ij}$ was discussed in ii)
and we concluded
that it doesn't correspond to a consistent triple. 
In the above two interrelated
facts we see the origin of the inconsistency in the FG expansion: 
the expansion should be
modified such that $\sigma B_{{\sss (2)}ij}$ disappears from the variation
and equivalently this should provide a prescription for 
the ``subtraction" in dimensional regularization.
The modification of the expansion proposed by FG \FG\ is to include 
logarithmic terms {\it i.e.} to replace \expansiong\ by:
\eqn\expansiongn{
g_{ij}(x,\rho)=\sum_{m=0}^\infty g_{{\sss (m)}ij}(x)\rho^m+ h_{{\sss (n)}ij}(x)
\rho^n\log(\rho)+...}
in $d=2n$ dimensions, where we made explicit just the first new term.
The PBH transformations 
\eqn\diffeo{\eqalign{
\rho&=\rho' e^{-2\sigma(x')}\simeq \rho'(1-2\sigma(x'))\,,\cr
x^i&=x'^i+a^i(x',\rho')\,,}}
induce now a modified transformation of $g_{{\sss (n)}ij}(x)$:
\eqn\modvar{
\bar \delta g_{{\sss (n)}ij}=\delta g_{{\sss (n)}ij}- 
2 \sigma h_{{\sss (n)}ij}\,,}
where $\delta g_{{\sss (n)}ij}$ is given in \var\ 
and give for the transformation of $h_{{\sss (n)}ij}$:
\eqn\hvar{\bar \delta h_{{\sss (n)}ij}= 2(1-n)\sigma h_{{\sss (n)}ij}\,.}
{}From \hvar\ it follows that  $h_{{\sss (n)}ij}$ 
should be one of the expressions
constructed under ii) 
above which transform homogeneously. 
In $d=4$, by choosing
therefore  $ h_{{\sss (2)}ij}={1\over2}B_{{\sss (2)}ij}$,  
due to the modified \modvar\
the unwanted term in the transformation $g_{{\sss (2)}ij}$ 
is cancelled and we have
now the consistent solution corresponding to \typeb.

We can easily follow through what happened in terms of a ``subtraction":
If we apply the operator ${1\over \sqrt g}{\delta \over \delta g^{ij}(x)}$
to \split\ we see that introducing the logarithmic term which eliminated the
second term in the variation is equivalent to ``subtracting" the second term
in \split. The solution of the cohomological problem originates now 
in the first term in  \split\ which engenders \typeb.
We believe that the two particular cases discussed in detail 
above from both points of 
view contain the basic features of the ``type A" and 
``type B" structures. Based
on that, we are ready to formulate some general rules related to the FG series:

a) In $d=2n $ there is a nontrivial cohomology in the Weyl transformation
of the $g_{{\sss (n)}ij}$ coefficient. As a
consequence, in the variation of the coefficient 
produced recursively by  the PBH
transformations there will be cohomologically nontrivial contributions.

b) These contributions are signaled in dimensional regularization by pole terms
in the solution: with vanishing residue in $d=2n$ for ``type A" and
nonvanishing residue for ``type B".

c) The contribution to $g_{{\sss (n)}ij}$
in $d=2n$ coming from ``type A" is nonlocal and it is
obtained by applying the ${1\over \sqrt g}{\delta \over \delta g^{ij}(x)}$
operator on the effective action which generates the Euler trace anomaly.

d) The contribution to $g_{{\sss (n)}ij}$ coming from ``type B",
whose number increases with the dimension, needs subtractions.
The coefficient of the logarithmic term  $h_{{\sss (n)}ij}$ in the expansion
is a linear combination of homogeneously transforming expressions which are
constructed again from ``type B" anomalies. 
There is a one-to-one correspondence between
these expressions and the subtractions.

e) After the subtractions are made, 
the contributions of ``type B" to $g_{{\sss (n)}ij}$
are nonlocal  and can be obtained by applying 
${1\over\sqrt g}{\delta\over\delta g^{ij}(x)}$ on
 the pieces of the effective 
action producing the various ``type B" trace anomalies.

We see that following 
the procedure above the so-called ``FG ambiguity"
is removed. One obtains  unique, well-defined, 
albeit generally nonlocal expressions
for $g_{{\sss (n)}ij}$. We remark  that due
 to the ``type B" contributions
an arbitrary scale $\mu$ will appear in the solution.
Of course one can still add to the solution terms transforming homogeneously
which have vanishing trace and covariant divergence.  

\newsec{Discussion and Conclusions}

In \Solo,\Eng\ the role of the boundary terms in the action was analysed.
These authors showed that the functional derivatives of the boundary action
are related to the FG coefficients. Therefore, using our
expressions for the FG coefficients, one can integrate the relation and obtain
explicitly the contribution of the boundary to the effective action.
{}From the discussion in Section 4 it is clear that the 
contributions will contain
$W(g)$  used in equations \act, \ggdef\ to generate the cohomologically 
nontrivial solutions. These terms are exactly the nonlocal terms generating
the anomalies. Their coefficients are fixed 
partially by the PBH transformations
and completely by the equations of motion. We have therefore identified 
the mechanism which produces the nonlocal anomalous terms in the effective
action. We stress that the calculation described above is done entirely
within the framework of classical gravity and does not require any knowledge
of the energy-momentum tensor of the dual CFT. In $d+1=3$ dimensions
an explicit calculation along these lines can be done both for the purely
bosonic and the supersymmetric cases \VANDRIEL. Indeed the Liouville
action and its supersymmetric generalizations appear in the effective action:
as it is well known the Liouville action represents the nonlocal
anomalous Polyakov action in the conformal gauge.

The translation of
an anomaly which is an eminently quantum effect into a classical
calculation is not entirely unexpected: in two dimensions  bosonisation
achieves this for the chiral anomaly. In the case discussed here this
is happening, however, for all dimensions  due to the AdS/CFT duality.  

Previously 
the coefficients of the anomaly were determined by looking at the 
infrared divergent terms in the bulk action. The above described procedure
gives an alternative way to calculate them, directly from the FG coefficients.
As discussed in Section 2, these two ways of calculating the anomaly
appear also on the CFT side of the equivalence. The relation  between the
two calculations is based on the Weyl invariance of the total effective action
in dimensional regularization. In the gravitational case probably
one can use directly diffeomorphisms in $d+1$ dimensions. We defer a detailed
discussion of the anomaly coefficients to a further publication.

An amusing feature is the appearance of a scale due to the type B anomalies.
The existence of an arbitrary (``ultraviolet'') scale in a classical
calculation is rather surprising.
 
As we mentioned above there is still some ambiguity left, related to 
nonlocal terms which are invariant under all the symmetries. We believe
these terms cannot be obtained by (anomalous) symmetry considerations
but they reflect genuine dynamical stringy corrections.

\bigskip
\noindent{\bf Acknowledgements:} We would like the 
Erwin-Schroedinger-Institute in Vienna, where most of this work was done, 
for hospitality. 

\bigskip

\noindent{\bf Appendix}

We discuss now in detail the proof that the Weyl variation 
$A_{{\sss (n)}ij}^E $ can be obtained as the inhomogeneous piece in the 
variation of a local tensor $\bar g_{ij}^E$.
We start by remarking that since the Weyl variation of the Euler density
$E_{2n}$, $\delta E_{2n}=-2n\sigma E_{2n}+{\rm total~derivative}$,
%
$$
\delta_\sigma\int d^d x\sqrt{g}E_{2n}=(d-2n)\int d^d x\sqrt{g}E_{2n}\sigma
\eqno({\rm A.1})
$$
where the curvature tensors entering $E_{2n}$ are in $d$ dimensions.  
Using (A.1) we can write $A_{{\sss (n)}ij}^E $ as 
%
$$
A^E_{{\sss(n)}ij}={1\over d-2n}{1\over\sqrt{g}}{\delta\over\delta g^{ij}}
\delta_\sigma\int d^d x\sqrt{g} E_{2n}
={1\over d-2n}\delta^{\rm (in)}_\sigma{1\over\sqrt{g}}
{\delta\over\delta g^{ij}}\int d^d x\sqrt{g}E_{2n}
\eqno({\rm A.2})
$$
where in the second step we used \com\ and ``$\delta^{({\rm in})}_{\sigma}$''
denotes the inhomogeneous part of the Weyl variation.
We have dropped a term which vanishes at $d=2n$.

The expression
%
$$
X_{{\sss(n)}ij}
\equiv{1\over\sqrt{g}}{\delta\over\delta g^{ij}}\int d^d x\sqrt{g}E_{2n}
\eqno({\rm A.3})
$$
vanishes in $d=2n$ when  the Euler density itself becomes a total derivative.
The ``topological identity'' this represents is:
%
$$
\delta^{[i_1}_{[j_1}\delta^{i_2}_{j_2}\cdots\delta^{i_{2n-1}]}_{j_{2n-1}]}
C_{[i i_{2n-1}}^{[j j_{2n-1}}C^{j_1 j_2}_{i_1 i_2}\cdots 
C^{j_{2n-3}j_{2n-2}]}_{i_{2n-3}i_{2n-2}]}
-{1\over 2n}\delta_i^j\delta^{[i_1}_{[j_1}\cdots\delta^{i_{2n-1}}_{j_{2n-1}}
\delta^{i_{2n}]}_{j_{2n}]}C_{[i_1 i_2}^{[j_1 j_2}\cdots
C_{i_{2n-1}i_{2n}]}^{j_{2n-1}j_{2n}]}=0
\eqno({\rm A.4})
$$
where $C^{ij}_{kl}$ is the Weyl tensor in $d=2n$.
In $d=2n$, $X_{{\sss(n)}ij}$ is proportional 
to the left hand side of (A.4).
Now we can identify a ``dimensionally continued''  $X_{{\sss(n)}ij}^{(d)}$
by replacing the $2n$ dimensional Weyl tensors in (A.4) with
the $d$ dimensional ones $C^{{\sss (d)}ij}_{kl} $ which transform homogeneously
in $d$ dimensions. The tensor $C^{{\sss (d)}ij}_{kl} $ is given explicitly
by:
%
$$
C^{(d)}_{kl}{}^{ij}=R_{kl}{}^{ij}-{1\over d-2}
\left(\delta_l^j R_k^i+\delta_k^i R_l^j
-\delta_l^i R_k^j-\delta_k^j R_l^i\right)
-{1\over(d-1)(d-2)}\left(\delta_l^i\delta_k^j-\delta_l^j\delta_k^i\right)\,.
\eqno({\rm A.5})
$$
Then we can expand the explicit $d$ dependence of $X_{{\sss (n)}ij}^{(d)}$
around $d=2n$, the curvature tensors continuing to be defined 
in $d$ dimensions:
%
$$
X_{{\sss(n)}ij}^{(d)}=X_{{\sss(n)}ij}-(d-2n)\bar g_{{\sss(n)}ij}^E\,,
\eqno({\rm A.6})
$$
where $\bar g_{{\sss(n)}ij}^E$ is explicitly defined as a local tensor 
by (A.6). Since  $X_{{\sss (n)}ij}^{(d)}$ transforms homogeneously, we have:
%
$$
{1\over d-2n}\delta_\sigma^{\rm (in)}X_{{\sss(n)}ij}
=\delta_\sigma^{\rm (in)}\bar g_{{\sss(n)}ij}^E
\eqno({\rm A.7})
$$
and therefore
%
$$
A_{{\sss (n)}ij}^E=\delta_\sigma^{\rm (in)}\bar g_{{\sss(n)}ij}^E\,.
\eqno({\rm A.8})
$$
We have thus shown that the variation 
of the functional derivative of the anomalous
part of the action related to the Euler type anomaly
can be integrated to a local functional. This
of course doesn't mean that the variation of the
action itself can be integrated to a local action:
taking the functional derivative breaks explicitly
the Bose symmetry between the different insertions
of the background metric, a symmetry which is a built-in 
requirement of the action itself.

{}From the above proof it is clear that in 
dimensional regularization the ``signature"
of the ``type A" nontrivial solution is the term ${X_{{\sss(n)}ij}\over d-2n}$.
In the spirit of the general mechanism discussed in Section 2, such a $ 0/0$
expression indicates the presence of a nonlocal, 
cohomologically nontrivial functional in integer dimension.
Calculating the Weyl variation, trace and covariant divergence of
${X_{{\sss(n)}ij}\over d-2n}$, we conclude 
that in $d=2n$ it corresponds to the triple \typea.
 
\listrefs
\end